\input{aipcheck}

\documentclass[
    ,final            
  ]
  {aipproc}

\layoutstyle{6x9}

\begin{document}

\title{Hyperaccreting Disks around Neutrons Stars and Magnetars for GRBs: Neutrino Annihilation and Strong Magnetic Fields}

\classification{97.10.Gz, 98.70.Rz} \keywords {accretion: accretion
disks --- gamma rays: bursts --- magnetic fields --- neutrinos ---
stars: neutron}

\author{Dong Zhang}{
  address={Department of Astronomy, The Ohio State University, 140W. 18th Ave.,
Columbus, OH, 43210, USA} }

\author{Z. G. Dai}{
  address={Department of Astronomy, Nanjing University, Nanjing, 210093, China}
}

\begin{abstract}
Hyperaccreting disks around neutron stars or magnetars cooled via
neutrino emission can be the potential central engine of GRBs. The
neutron-star disk can cool more efficiently, produce much higher
neutrino luminosity and neutrino annihilation luminosity than its
black hole counterpart with the same accretion rate. The neutron
star surface boundary layer could increase the annihilation
luminosity as well. An ultra relativistic jet via neutrino
annihilation can be produced along the stellar poles. Moreover, we
investigate the effects of strong fields on the disks around
magnetars. In general, stronger fields give higher disk densities,
pressures, temperatures and neutrino luminosity; the neutrino
annihilation mechanism and the magnetically-driven pulsar wind which
extracts the stellar rotational energy can work together to generate
and feed an even stronger ultra-relativistic jet along the stellar
magnetic poles.
\end{abstract}

\maketitle


\section{Introduction}
The hyperaccreting disk surrounding a stellar-mass black hole
possibly formed by the merger of a compact object binary or the
collapse of a massive star has been argued to be a candidate for
central engine of gamma-ray bursts (GRBs; e.g., Popham et al. 1999).
On the other hand, both observational and theoretical evidences show
that newborn neutron stars or magnetars rather than black holes can
form in the GRB central engines (Usov 1992; Dai \& Lu 1998).
However, all the previous magnetized neutron star or magnetar models
ignore the accretion process which occurs onto a protoneutron star
for the first several seconds. In the collapsar scenario, if the
rotational core collapse can lead to the formation of a neutron star
or a magnetar, it is possible that the prompt accretion or fallback
process can make a hyperaccreting disk around the young formed star.
Also a debris disk around a massive neutron star can be formed in
compact binary mergers. Therefore we investigate the hyperaccreting
neutron star or magnetar system, and show that it can also proposed
as another possible central engine of GRBs (Zhang \& Dai 2008, 2009,
2010). For simplicity we consider the vertically integrated disk
using the $\alpha$-prescription. To our knowledge, our work is the
first to study the hyperaccreting disks around neutron stars
possibly related to GRBs.

\section{Neutrino-Cooled Disks around Neutron Stars}

In the hyperaccreting disk the Alfv\'{e}n radius of the central star
can be estimated as
$r_{A}\simeq0.207\dot{M}_{-2}^{-2/7}M_{1.4}^{-1/7} \mu_{30}^{4/7}
\textrm{km}$, where $\dot{M}=0.01\dot{M}_{-2}M_{\odot}$ s$^{-1}$ is
the accretion rate, $M=1.4M_{1.4}M_{\odot}$ is the central star
mass, $\mu=\mu_{30}10^{30}$ G cm$^{3}$ is the central magnetic flux.
If the surface field $B_{0}$ is less than $B_{0}\leq B_{\rm
crit}=0.89\times10^{15}\dot{M}_{-2}^{1/2}M_{1.4}^{1/4}
r_{*,6}^{-5/4}$ G with $r=r_{*,6}10^{6}$ cm being the star radius,
the accretion flow will continue to be confined in the disk plane
without co-rotating with the compact object or getting funneled onto
the magnetar poles.

In the weak field case, the quasi-steady disk around a neutron star
can be approximately divided into two regions --- inner and outer
disks, depending on the energy transfer and emission in the disk.
For the outer disk, the heating energy rate $Q^{+}$ is mainly due to
local dissipation ($Q^{+}=Q_{\rm vis}^{+}$), and the structure of
the outer disk is very similar to the black hole disk. On the other
hand, the heating energy in the inner disk includes both the energy
generated by itself and the energy advected from the outer region
($Q^{+}=Q_{\rm vis}^{+}+Q_{\rm adv}^{+}$), so the inner disk has to
be dense with a high pressure. We take $Q^{+}=Q^{-}$ and the
entropy-conservation self-similar condition to describe the inner
disk. The left panel of Figure 1 shows the size of the inner disk,
which is determined by the global energy equation of the inner disk.
Moreover, we also calculate the size of a low-$\alpha$ inner disk,
which is smaller for a low accretion rate ($\leq 0.1 M_{\odot}$
s$^{-1}$) compared to a high-$\alpha$ disk ($\alpha\sim 0.1$), and
increases dramatically with increasing accretion rate. Also, the
inner disk structure would not exist if the disk is unstable to
drive a large scale thermal non-relativistic wind when the accretion
rate $\geq0.5M_{\odot}$ s$^{-1}$.

We find that, due to the inner disk structure as well as the inner
surface boundary of the compact star, the disk has a denser, hotter
inner region with higher pressure compared to its black hold
counterpart. Also, the entire disk can cool more efficiently via
neutrino emission. The right panel of Figure 1 shows the total
neutrino emission luminosity of the entire disk around a neutron
star as a function of accretion rate, and we compare it with the
neutrino luminosity from a black-hole disk. We see that the
difference in neutrino luminosity between the neutron-star and
black-hole cases is a strong function of the accretion rate. When
the accretion rate is low, the total neutrino luminosity of the
black-hole disk $L_{\nu,\rm BH}$ is much smaller than that of the
neutron-star disk $L_{\nu,\rm NS}$, but $L_{\nu,\rm BH}$ and
$L_{\nu,\rm NS}$ are similar for a moderate accretion rate.

\begin{figure}
\centering\resizebox{0.8\textwidth}{!} {\includegraphics{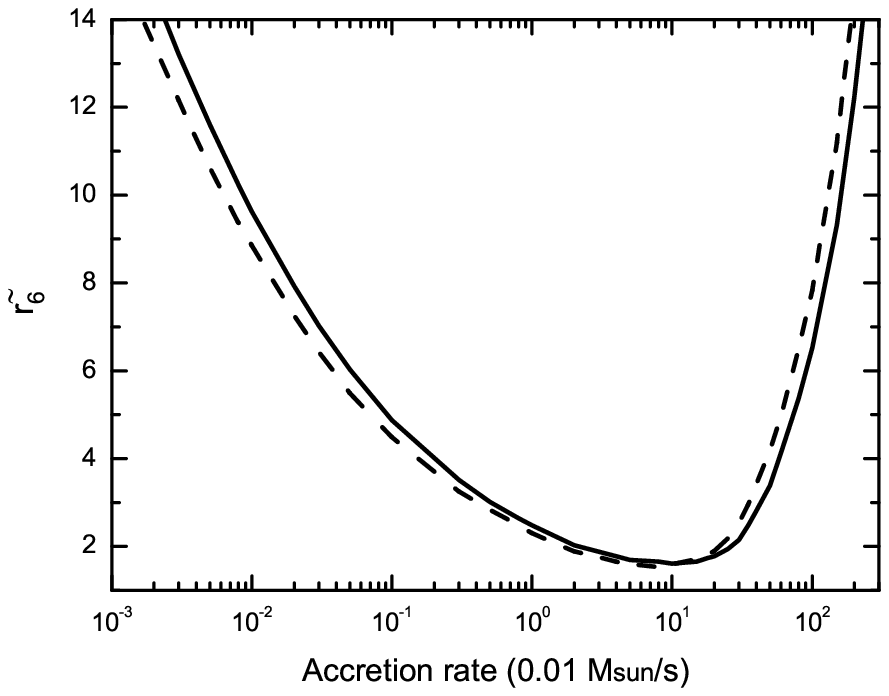}
\includegraphics{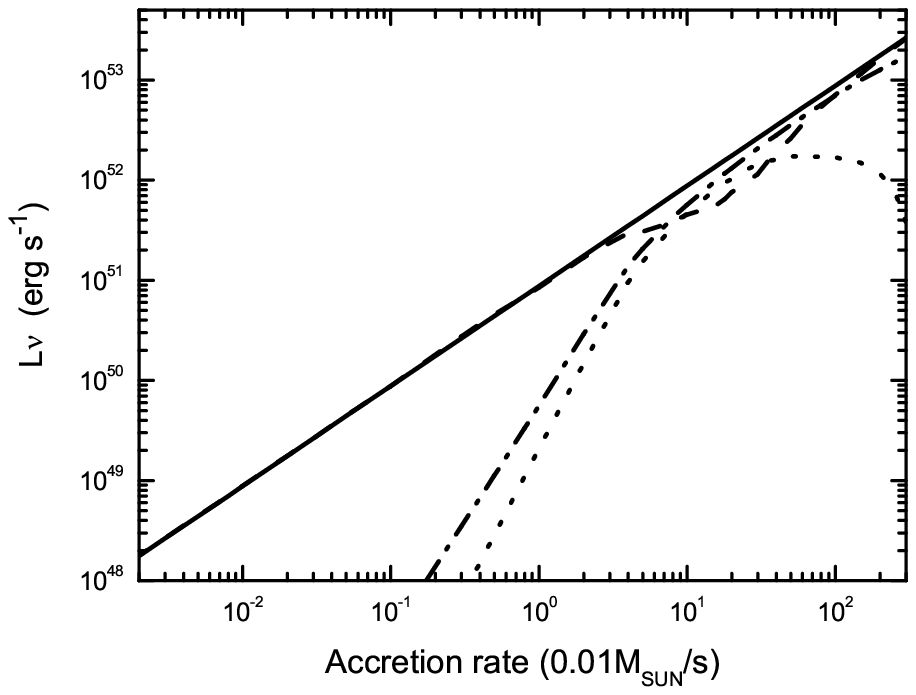}}
\caption{(a) {\em Left panel}: the radius $r/(10^{6}$cm) between
the inner and the outer disks with the center neutron star
$M=1.4M_{\odot}$ ({\em solid line}), and $M=2.0M_{\odot}$ ({\em
dashed line}). (b) {\em Right panel}: Neutrino luminosity from the disk with $M=1.4M_{\odot}$.
The solid line corresponds to the maximum energy release rate of the disk around a
neutron star, the dashed line to the neutrino luminosity from the
inner disk, the dotted line to the neutrino luminosity from the
outer disk, and the the dash-dotted line to the neutrino luminosity
from a black hole disk.}\label{fig1}
\end{figure}

\section{Neutrino Annihilation}

Hyperaccreting black hole disks can convert some fraction of the net
accretion energy into the energy of a relativistic jet by two
general mechanisms: neutrino annihilation and magnetohydrodynamical
(MHD) effects such as the Blandford-Znajek mechanism. However, for
hyperaccretion disks surrounding neutron stars, the energy
conversion mechanism is mainly due to the neutrino annihilation for
the magnetic fields $\leq10^{15}$ G. As the neutron star disk has a
brighter neutrino luminosity compared with the black hole disk, the
neutrino annihilation efficiency of the neutron star disk should be
higher than that of the black hole disk. Moreover, the surface
boundary of the neutron star, which carries away
gravitational-binding energy by neutrino emission makes the neutrino
annihilation luminosity of the neutron star disk be even higher. We
follow the approximate method used by Popham et al. (1999) to
calculate the annihilation rate, and we introduce the efficiency
factor $\eta_{s}$ to measure the energy emitting from the stellar
surface as $L_{s}\simeq\eta_{s}GM\dot{M}/(4r_{*})$ with $r_{*}$
being the neutron star radius.
\begin{figure}
\centering\resizebox{0.8\textwidth}{!} {\includegraphics{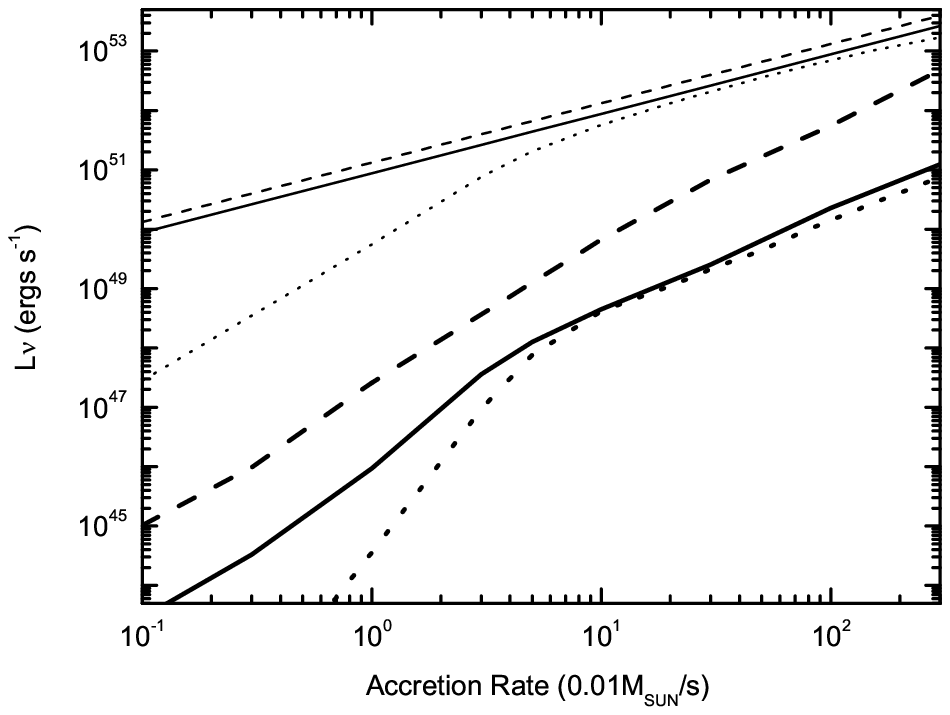}\includegraphics{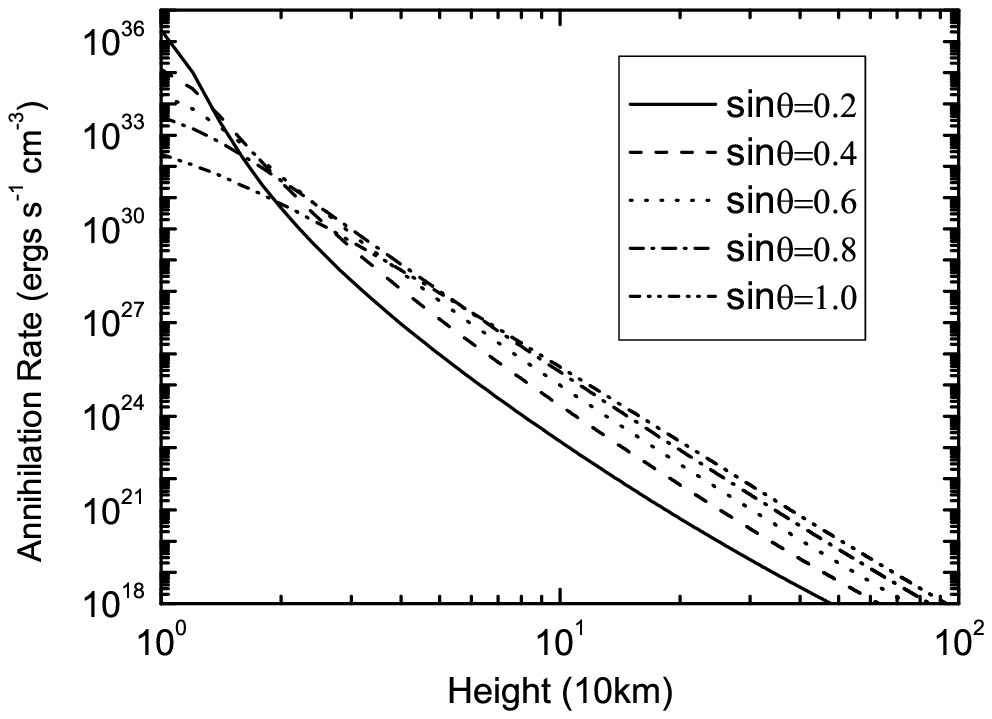}}
\caption{(a) {\em Left panel}: Neutrino annihilation luminosity $L_{\nu\bar{\nu}}$ ({\em thick lines}) and
total neutrino emission luminosity $L_{\nu}$ ({\em thin lines}) as
functions of accretion rate. The solid lines correspond to the
neutron star disk with the boundary layer emission efficiency $\eta_{s}=0$, the dashed lines to $\eta_{s}=0.5$,
and the dotted lines to the black hole disk. (b) {\em Right panel}: The
neutrino annihilation rate $l_{\nu\bar{\nu}}$  along the stellar magnetic pole as a function of height for
accretion rate $\dot{M}=0.5M_{\odot}$ s$^{-1}$ in strong field case,
where $\textrm{sin}\theta\simeq\sqrt{r_{*}/r_{A}}$ measures the strength of funnel accretion,
smaller $\textrm{sin}\theta$ corresponds to stronger field and funnel accretion.}\label{fig2}
\end{figure}

The left panel of Figure 2 shows the total neutrino annihilation
luminosity and the emission luminosity of a neutron-star disk with
different surface boundary layer conditions ($\eta_{s}$=0 and 0.5).
The annihilation luminosity of the neutron-star disk are brighter
than those of a black hole disk with the same mass and accretion
rate. If we study the neutrino annihilation from the entire disk
without surface boundary emission ($\eta_{s}$=0), the difference
between $L_{\nu\bar{\nu},NS}$ and $L_{\nu\bar{\nu},BH}$ is more
significant for a low accretion rate than for a high accretion rate.
On the other hand, neutrino emission from the neutron star surface
boundary layer ($\eta_{s}$=0.5) makes the annihilation luminosity be
more than one order of magnitude higher than that without boundary
emission ($\eta_{s}$=0). Therefore, a lower-spin neutron star
(larger $\eta_{s}$) with hyperaccreting disk around it could have an
obviously higher annihilation efficiency than that of a higher-spin
neutron star.

\section{Effects of Strong Magnetic Fields}

Strong fields $\geq10^{15}$ G from the center magnetar can play a
significant role in affecting the disk properties and even changing
the accretion process (Zhang \& Dai 2010). We consider the magnetar
field has a dipolar vertical component $B_{z}$. The differential
rotation between the disk and the magnetar will generate a toroidal
field component $B_{\phi}$ and a relatively weak radial component
$B_{r}$. The generated field can have an open or closed
configuration, depending on the disk's viscous turbulence, magnetic
diffusivity and disk angular velocity. For the strong field
environment, the quantum effects (Landau levels) and large scale
field coupling play two competitive roles in changing the disk
properties, the former to decrease the pressure, density and
neutrino luminosity with increasing field strength, while the latter
to increase them. However, in most cases the field coupling is more
significant than the microphysical quantum effect. Note that in the
hyperaccreting disks, the funnel accretion can only be important for
extremely strong fields. The funnel flow will cover a ring-like belt
of ``hot spot'' around the magnetar surface and emit thermal
neutrinos. Because of the more concentrated emission from the ``hot
spot", the funnel accretion can accumulate even more powerful
neutrino annihilation luminosity than the weak field case (see right
panel of Figure 2).

The neutrino annihilation process both from the magnetar surface and
from the disk plane will be higher than that without fields.
Moreover, the neutrino annihilation mechanism and the magnetic
activity from the stellar surface (i.e., the pulsar wind mechanism)
can work together to generate and feed an ultra-relativistic jet
along the stellar magnetic poles. If the stellar spin period is
sufficiently short (e.g., $\sim4$ ms for the field $\sim10^{16}$ G
and $\dot{M}=0.1M_{\odot}$ s$^{-1}$), the jet from the magnetar will
be magnetically-dominated and mainly feeded by extraction the
stellar rotational energy. If the magnetar spin period is long, the
jet is thermally-driven and feeded by the annihilation process. In
this case, an energetic ultrarelativistic jet via neutrino
annihilation can be produced above the stellar polar region if the
disk accretion rate and the stellar surface luminosity are
sufficiently high. In the intermediate case, on the other hand, the
relativistic jet can be launched by the pulsar-wind-like process and
neutrino annihilation together.





\bibliographystyle{aipproc}   

\bibliography{sample}

\IfFileExists{\jobname.bbl}{}
 {\typeout{}
  \typeout{******************************************}
  \typeout{** Please run "bibtex \jobname" to optain}
  \typeout{** the bibliography and then re-run LaTeX}
  \typeout{** twice to fix the references!}
  \typeout{******************************************}
  \typeout{}
 }

\end{document}